\documentclass[twocolumn,aps,floatfix,pra]{revtex4}
\usepackage{amsmath}
\usepackage{graphicx}
\usepackage[suffix=]{epstopdf}
 
\begin{document}

\title{Momentum correlations of a few ultra-cold bosons \\escaping from an open well}
\author{Jacek Dobrzyniecki}
\email{Jacek.Dobrzyniecki@ifpan.edu.pl}
\author{Tomasz Sowi\'nski}
\affiliation{Institute of Physics, Polish Academy of Sciences, Aleja Lotnikow 32/46, PL-02668 Warsaw, Poland}

\begin{abstract}
The dynamical properties of a one-dimensional system of two and three bosons escaping from an open potential well are studied in terms of the momentum distributions of particles. In the case of a two-boson system, it is shown that the single- and two-particle momentum distributions undergo a specific transition as the interaction strength is tuned through the point where tunneling switches from the pair tunneling to the sequential one. Characteristic features in the momentum spectra can be used to quantitatively determine the participation of specific decay processes. A corresponding analysis is also performed for the three-boson system, showing a scheme for generalizations to higher particle numbers. For completeness, the time-dependent Tan's contact of the system is also examined and its dynamics is found to undergo a similar transition. The results provide insight into the dynamics of decaying few-body systems and offer potential interest for experimental research. 
\end{abstract}

\maketitle 
\section{Introduction}

The quantum tunneling through a classically impenetrable barrier is one of the oldest problems in quantum mechanics. This problem arises in the analysis of such phenomena as the $\alpha$-decay of an atomic nucleus \cite{gamow1928}, proton emission \cite{talou1999,talou2000}, fusion, fission, photoassociation, and photodissociation \cite{keller1984,bhandari1991,balatenkin1998,vatasescu2000}. While the tunneling of a single particle is well understood \cite{razavy2003}, and the tunneling of a Bose-Einstein condensate of a large number of particles is well described by a mean-field approximation \cite{salasnich2001,carr2005,huhtamaki2007,zhao2017}, the escape behavior of interacting few-body systems is a more complicated problem in which the dynamics is nontrivially governed by the interplay between interactions, indistinguishability and quantum correlations \cite{gogolin2004}. Although significant attention has been already devoted to the dynamical processes of a few particles confined in closed lattice potentials \cite{winkler2006,folling2007,zollner2008a,chen2011}, in recent years the significantly different problem of a few particles tunneling into open space has attracted an increasing amount of interest \cite{delcampo2006,lode2009,kim2011,taniguchi2011,lode2012,maruyama2012,hunn2013,lode2014,maksimov2014,gharashi2015,lundmark2015,lode2015,dobrzyniecki2018}.

Current experimental advancements in the field of ultra-cold physics allow realizing a variety of formerly purely theoretical scenarios in the lab \cite{blume2012,zinner2016,sowinski2019}. In particular, it is possible to precisely control the shape of the external confinement \cite{meyrath2005,henderson2009,vanes2010}, the inter-particle interactions \cite{pethick2008,zollner2008b,chin2010}, as well as engineer the initial state of the system \cite{pethick2008,serwane2011} and mimic low-dimensional physics \cite{gorlitz2001,greiner2001,schreck2001}. The recent experiments on the tunneling of a few interacting atoms escaping from an effectively one-dimensional potential well \cite{zurn2012,zurn2013} provide fresh motivation for the study of such tunneling problems. Since the presence of inter-particle interactions and quantum correlations affects the tunneling of few-body systems in interesting and complex ways, a deeper understanding of this issue could become important also from a theoretical point of view \cite{delcampo2006,lode2009,kim2011,taniguchi2011,lode2012,maruyama2012,hunn2013,lode2014,maksimov2014,gharashi2015,lundmark2015,lode2015,dobrzyniecki2018}.

Although the system of two tunneling particles is the simplest possible case of the few-body tunneling problem, many open questions still remain unanswered, and it continues to be explored in recent research \cite{ishmukhamedov2017,ishmukhamedov2018}. In \cite{lode2012,lode2014} the properties of the system were partially studied from the momentum distribution point of view for repulsive interactions. Here we extend this description by also taking the attractive branch of interactions into account. This allows us to examine how the momenta of the system are changed when the interaction strength is tuned across the point of transition between sequential tunneling (when bosons leave the well one by one) and collective tunneling (when bosons leave the well as clusters of two or more particles) \cite{rontani2012,rontani2013,dobrzyniecki2018}. Precise relationships can be established that connect the form of the momentum spectra to quantities such as the system energy and the relative participation of the different tunneling processes. Apart from the two-boson system, we also touch upon the more complicated three-boson case, showing how the analysis may be generalized to higher particle numbers. In addition, we touch upon the time evolution of the Tan's contact, a quantity related to the interaction energy between the particles \cite{tan2008,tan2008a,tan2008b}. 

The work is organized as follows. In Section \ref{sec:model} we describe the model system under study. In Section \ref{sec:eigenspectrum} we examine the eigenspectrum of the many-body Hamiltonian of the open well system. In Section \ref{sec:dynamics} we describe the decay dynamics of a two-boson system, showing the basic nature of the tunneling dynamics, and the transition between distinct regimes that occurs at a specific value of the interaction strength. In Section \ref{sec:momentum} we discuss the momentum distribution of the decaying two-boson system. In Section \ref{sec:transition} we focus on the ways in which the transition between different regimes is reflected in the center-of-mass momentum distribution and the Tan's contact. In Section \ref{sec:3bosons} we discuss the momentum distributions of a three-boson system. Section \ref{sec:conclusion} is the conclusion. 

\section{The model}
\label{sec:model}

In this work we consider a system of $N=2$ and $N=3$ indistinguishable ultra-cold bosons of mass $m$, interacting via a point-like $\delta$ potential and confined in an effectively one-dimensional external trap. The many-body Hamiltonian of this system has the form: 
\begin{equation}
\label{mb_hamiltonian_1}
    H = \sum\limits_i \left[ -\frac{\hbar^2}{2m} \frac{\partial^2}{\partial x_i^2} + V(x_i) \right] + g \sum\limits_{i<j} \delta(x_i-x_j),
\end{equation}
where $x_i$ represents the position of the $i$-th boson and $V(x)$ is the external potential. The effective inter-particle interaction strength $g$ is related to the three-dimensional $s$-wave scattering length \cite{olshanii1998,haller2009} and its magnitude can be tuned experimentally via the Feshbach resonances \cite{pethick2008,chin2010} or by changing the confinement in perpendicular directions \cite{olshanii1998}. Recently it was argued that a potential boson species feasible for experimental realization of such a system is $^{87}\mathrm{Rb}$ or $^{85}\mathrm{Rb}$ atoms \cite{delcampo2006}. It is worth mentioning that few-body systems of bosonic $^{87}\mathrm{Rb}$ atoms have already been prepared in optical lattices \cite{folling2007,trotzky2008,bakr2010,sherson2010}.

\begin{figure}
\includegraphics[width=1\linewidth]{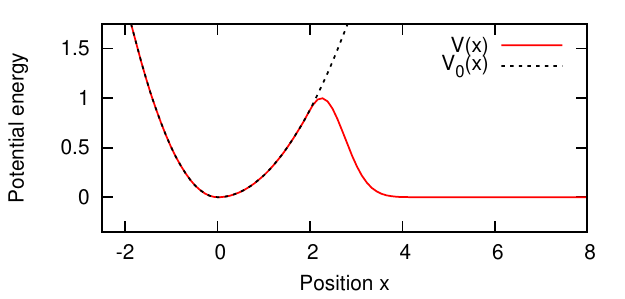}
\caption{The shape of the trapping potential at $t < 0$ ($V_0(x)$, black dashed line) and after a sudden change at $t \ge 0$ ($V(x)$, solid red line). Energy and length are given in units of $\hbar\Omega_0$ and $\sqrt{\hbar/m\Omega_0}$, respectively.}
 \label{Fig1} 
\end{figure}

\begin{figure}
\includegraphics[width=1\linewidth]{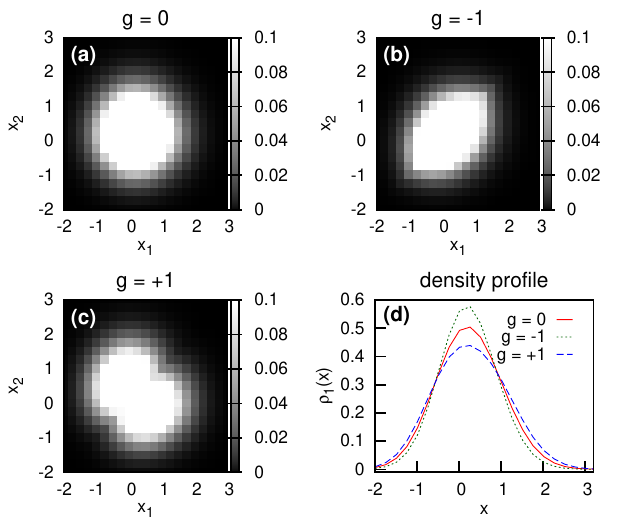}
\caption{The two-particle probability density $\rho_2(x_1,x_2)$ of the initial state for the two-particle system with zero interactions $g = 0.0$ (top left), attractive interactions $g = -1.0$ (top right) and repulsive interactions $g = 1.0$ (bottom left). Bottom right: The single-particle density profile $\rho_1(x)$ of the initial state for different interactions. It can be seen that repulsive interactions broaden the single-particle density profile and reduce the two-particle density along the $x_1 = x_2$ diagonal, while attractive interactions narrow the single-particle density profile and concentrate two-particle density along the diagonal. Positions are in units of $\sqrt{\hbar/m\Omega_0}$, interaction strength in units of $\sqrt{\hbar^3\Omega_0/m}$.}
 \label{Fig2} 
\end{figure}

We assume that at $t < 0$ bosons are confined within a closed asymmetric well potential $V_0(x)$ (dashed line in Fig.~\ref{Fig1}): 
\begin{equation}
 V_0(x) = 
 \begin{cases}
    m\Omega_0 x^2/2 ,& x< 0 \\   
    m\Omega x^2/2 ,& x>0, 
\end{cases}
\end{equation}
where the modified frequency $\Omega\approx\Omega_0/2.26$. Accordingly, the initial many-body state at $t = 0$ is chosen as the ground state of an interacting $N$-boson system confined inside $V_0(x)$. For given interaction $g$, we find the ground state numerically by propagating in imaginary time a trial many-body wave function (chosen as the ground state of $N$ non-interacting bosons in a harmonic oscillator well). In Fig.~\ref{Fig2}, we show the single- and two-particle density profiles ($\rho_1(x) = \int \mathrm{d}x_2 |\Psi(x,x_2)|^2$ and $\rho_2(x_1,x_2) = |\Psi(x_1,x_2)|^2$) of the obtained initial state for $N=2$ bosons for three generic interaction strengths $g$: non-interacting ($g = 0$), repulsive ($g = 1$) and attractive ($g = -1$). In the initial state the bosons are completely contained within the well region. For the non-interacting case (Fig.~\ref{Fig2}a), the two-particle density $\rho_2(x_1,x_2)$ corresponds to a product of two identical single-particle wave functions that have a nearly Gaussian profile reflected by $\rho_1(x)$ (solid red in Fig.~\ref{Fig2}d). In the presence of attractive (repulsive) interactions, the bosons are more likely (less likely) to be near each other and the density $\rho_2(x_1,x_2)$ is concentrated closer to (away from) the $x_1=x_2$ diagonal (Fig.~\ref{Fig2}b,\ref{Fig2}c). It is reflected in the shrinking (broadening) of the single-particle density profile $\rho_1(x)$ (dotted green and dashed blue in Fig.~\ref{Fig2}d).

At $t = 0$ the trap is suddenly opened and the external potential is changed to $V(x)$, given by the following expression ($x_0 = \sqrt{\hbar/m\Omega_0}$ is the harmonic oscillator length unit): 
\begin{equation}
 V(x) = 
 \begin{cases}
    m\Omega_0 x^2/2 ,& x< 0 \\   
    m\Omega x^2/2 ,& 0 \le x\le 2x_0\\   
  \hbar\Omega_0 \mathrm{e}^{-2(x/x_0-9/4)^2} ,& x > 2x_0.
\end{cases}
\end{equation}
The resulting shape of $V(x)$ is that of a potential well separated from open space by a finite barrier (solid line in Fig.~\ref{Fig1}.) The function $V(x)$ is chosen so that both it, and its first derivative, are continuous everywhere. 

The time evolution of the many-body state for $t > 0$ is obtained straightforwardly by solving numerically the time-dependent many-body Schr\"odinger equation in position representation. The equation is integrated using a fourth-order Runge-Kutta method. The calculations are done on a dense grid (with spacing $\delta x = 0.25x_0$ for the two-boson case and $\delta x = 0.33x_0$ for the three-boson case) in a region that includes the potential well and a large extent of space outside \cite{dobrzyniecki2018}. The boundaries of the simulated region are chosen large enough to ensure that reflections from the hard-wall boundaries do not affect the main results. For the $N=2$ case the simulated region is given by $x \in [-10x_0,+200x_0]$, while for the $N=3$ case it is given by  $x \in [-4x_0,+120x_0]$. In contrast to our previous study \cite{dobrzyniecki2018}, we do not use a complex absorbing potential at the boundaries, since then, as argued in \cite{selsto2010}, it would be impossible to properly study correlations between particles inside and outside the well.

For convenience, in all further discussion we use harmonic oscillator units, {\it i.e.} energy, length, and the interaction strength are given in $\hbar \Omega_0$, $\sqrt{\hbar/m\Omega_0}$, and $\sqrt{\hbar^3\Omega_0/m}$, respectively. 

\section{The Hamiltonian eigenspectrum}
\label{sec:eigenspectrum}

\begin{figure}
\includegraphics[width=1\linewidth]{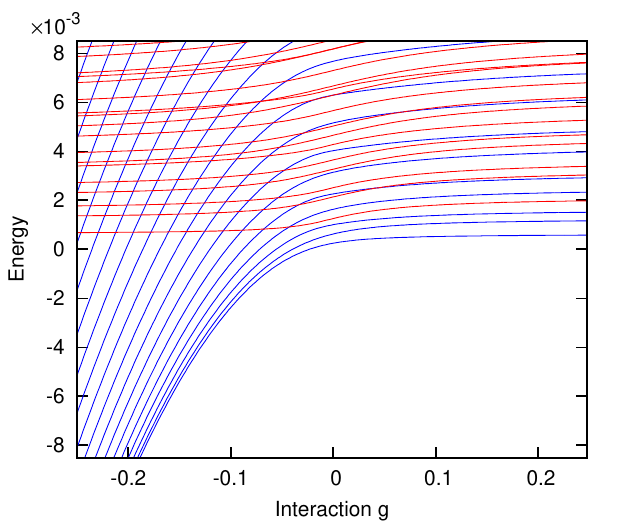}
\caption{The eigenenergies of the many-body Hamiltonian \eqref{mb_hamiltonian_1} after the well is opened at $t = 0$, for $N = 2$ bosons and different interaction strength $g$. The eigenstates can be divided into two groups: states which describe a pair of almost independent particles for all $g$ (red), and states which for $g > 0$ describe independent particles but for $g < 0$ describe bound boson pairs with energy strongly dependent on $g$ (blue). Energies are in units of $\hbar\Omega_0$, interaction strength in units of $\sqrt{\hbar^3\Omega_0/m}$.}
 \label{Fig3} 
\end{figure}

To build some intuition of the system properties, we first examine the spectrum of the many-body Hamiltonian \eqref{mb_hamiltonian_1} after the trap is opened at $t = 0$ for a system with $N = 2$ bosons (Fig.~\ref{Fig3}). It is very useful to refer this spectrum to the well-known eigenspectrum of the Gaudin-Lieb-Liniger model, describing a one-dimensional system of $N$ ultracold bosons in free space, adapted to the $N = 2$ case \cite{lieb1963a,lieb1963b,gaudin1971,batchelor2005,takahashi2005}. Although the system studied is slightly different from the original model (nontrivial confinement in the initial region), the eigenspectra of both Hamiltonians are very similar and all eigenstates can be divided into two well-distinguished groups. The first group consists of states of almost independent particles having momenta $k_1$ and $k_2$, \emph{i.e.}, the total energy of the state is $E \approx k_1^2/2 + k_2^2/2$. Since this energy is almost independent of interaction strength $g$, these states are represented by nearly horizontal lines in the plot (shown in red). The second group of states represents particles which become bounded for attractive forces. It means that for $g < 0$ the bounded pair has a total energy directly dependent on interactions, $E \approx K^2/4 - g^2/4$, where $K$ is the center-of-mass momentum \cite{lieb1963b}. For repulsive forces ($g > 0$) these states behave similarly to states from the first group and their energy is nearly independent of interactions. Consequently, these states are represented in Fig.~\ref{Fig3} by characteristic parabolic lines on the attractive branch which smoothly evolve into horizontal lines on the repulsive branch (shown in blue). 

Similar argumentation can be also applied to the system of $N = 3$ bosons. By comparison to the Gaudin-Lieb-Liniger model, in this case we expect three groups of many-body eigenstates \cite{takahashi2005}. The first group includes states with three almost independent particles with energy independent of interactions, $E \approx k_1^2/2 + k_2^2/2 + k_3^2/2$. The second group contains states of three particles which for $g < 0$ form a composition of a bound pair and a third independent particle. The corresponding eigenenergy can be written as $E \approx K^2/4 - g^2/4 + k_3^2/2$, where $K$ is the center-of-mass momentum of the pair. Finally, the third group is built from states which for $g < 0$ describe bound trimers with energies $E \approx P^2/6 - g^2$, where $P$ is the center-of-mass momentum of the trimer. 

The evolution of an initial state prepared as the interacting ground state of the confined system depends on its direct decomposition into many-body eigenstates of the post-quench Hamiltonian (the eigenspectrum itself does not directly provide information about which of the different tunneling mechanisms will dominate in the dynamics). Therefore, tunneling processes other than sequential tunneling, which involve clusters of two or more particles, become available only for $g < 0$. Unfortunately, due to numerical complexity, the aforementioned decomposition cannot be performed efficiently and accurately, and one needs to use other approaches to answer the question of the participation of different tunneling processes. Having in mind that different tunneling processes are associated with specific momentum correlations, in the following we address this question by performing a numerically exact time evolution of the system and analyzing the momentum distributions. 

\section{Dynamics of two escaping bosons}
\label{sec:dynamics}

After the well is suddenly opened at $t = 0$, the bosons start escaping through the barrier into open space. Before we analyze the two-boson system from the momentum distribution point of view, let us recall recent results from \cite{dobrzyniecki2018} and shortly discuss the dynamical properties of the system from the point of view of density distributions. It is known that the dynamical properties of the system depend significantly on the strength of inter-particle interactions. As the interaction strength is changed from repulsions to sufficiently strong attractions, the dynamics of the two-boson system undergoes a transition between two distinct scenarios, characterized by the dominance of different decay processes. Below a critical value of interactions (approximately $g = -0.9$ in the case studied), essentially the entire decay is dominated by pair tunneling, {\it i.e.}, both bosons leave the well simultaneously as a bound pair. Conversely, for $g$ above the critical value, the decay is dominated by a sequential tunneling, in which the bosons leave the well one by one. In such a regime, pair tunneling can still occur but is significantly less likely. In fact, for $g \ge 0$, for which a two-boson bound state does not exist, pair tunneling essentially vanishes.

\begin{figure}
\includegraphics[width=1\linewidth]{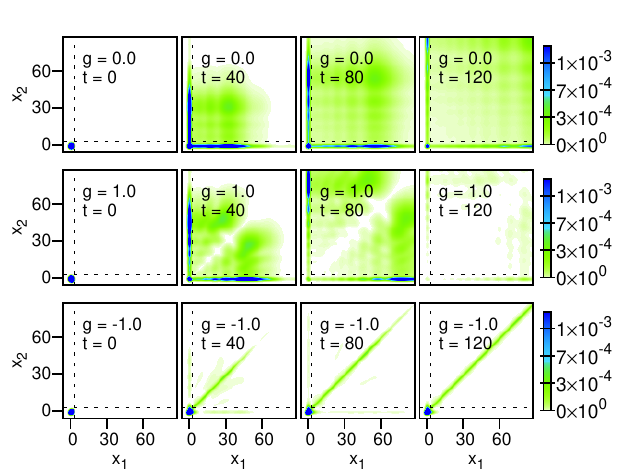}
\caption{Time evolution of the two-particle probability density $\rho_2(x_1,x_2;t)$ of an initially trapped two-boson system for various interaction strengths $g$. The dashed lines demarcate the well boundary $x_B = 3x_0$. For the non-interacting and repulsive systems ($g = 0.0, g = 1.0$), essentially the entire decay process takes place via sequential tunneling, while for sufficiently strong attractions ($g = -1.0$) the system decays almost solely via pair tunneling. Positions are in units of $\sqrt{\hbar/m\Omega_0}$, interaction strength in units of $\sqrt{\hbar^3\Omega_0/m}$, time in units of $1/\Omega_0$. Compare also to Fig. 2 in \cite{dobrzyniecki2018}.}
 \label{Fig4} 
\end{figure}

To illustrate the transition between the two tunneling mechanisms, in Fig.~\ref{Fig4} (exactly as in \cite{dobrzyniecki2018}) we show snapshots of the two-particle density profile $\rho_2(x_1,x_2;t) = |\Psi(x_1,x_2;t)|^2$ for different moments after the opening of the well, and different interaction strengths $g$. For better visibility we indicate the well boundary $x_B\approx 3x_0$ with dashed lines. 
 
In the most trivial scenario of vanishing interactions ($g = 0$) both bosons tunnel entirely independently of each other. After a short time ($t=40$) a significant amount of the density is present in the region $(x_1-x_B)(x_2-x_B)<0$ indicating a high probability of exactly one boson being outside the well. For longer times both bosons are likely to end up outside the well. Due to the absence of interactions, the two-particle wave function is simply a product of two identical single-particle wave functions during the entire process.

For a repulsive system ($g = 1.0$) the dynamics is also dominated by the sequential tunneling, indicated by the initial accumulation of the probability density in $(x_1-x_B)(x_2-x_B)<0$ region. However, the inter-particle repulsion causes a significant anticorrelation in the boson positions, so that the probability density vanishes close to the $x_1 = x_2$ diagonal. It is clear that simultaneous tunneling of bosons is strongly suppressed in this case. 

For sufficiently strong attractions the scenario is completely different (see example results for $g=-1.0$ in Fig.~\ref{Fig4}). The sequential tunneling is suppressed and the bosons leave the well only as a composite pair. This can be seen from the density distribution, which during the whole evolution is nonzero only in the regions with  $(x_1-x_B)(x_2-x_B)>0$ and remains concentrated around the line $x_1 = x_2$. 

It is useful to compare the described pair tunneling mechanism with an analogous process in an optical lattice, recently observed  \cite{winkler2006,folling2007,zollner2008a,chen2011,dobrzyniecki2016}. In the latter case, the co-tunneling of an atom pair between neighboring sites is forced by the energy conservation: two interacting particles occupying the same site are forced to tunnel as a pair to avoid energy mismatch. This mechanism is present for attractive as well as repulsive interactions. By contrast, collective tunneling from a single well to the open space (as shown above) occurs only when particles form a bound state, which in a one-dimensional scenario is possible only for attractive interactions.

\section{Momentum distributions}
\label{sec:momentum}
Now let us discuss how the different tunneling regimes are reflected in the particle momenta. For this purpose we study the time evolution of the two- and single-particle momentum distributions defined as
\begin{subequations} \label{MomentumDef}
\begin{align}
\pi_2(k_1,k_2;t) &= \frac{1}{4\pi^2\hbar^2} \left|\int\!\mathrm{d}x_1\mathrm{d}x_2\,\mathrm{e}^{-i(k_1 x_1+k_2x_2)/\hbar}\Psi(x_1,x_2;t)\right|^2, \\
\pi_1(k;t)&=\int\!\mathrm{d}k'\,\pi_2(k,k';t).
\end{align}
\end{subequations}
It should be pointed out that, from the experimental point of view, measuring momentum distributions is quite feasible, since appropriate techniques have been developed for measuring positions and velocities \cite{she1978,pan1980,prodan1982,mabuchi1996,ottl2005,bondo2006,heine2010} of individual untrapped atoms. For the specific problem of bosons escaping from a potential well, a relevant experimental scheme to measure the momenta of the emitted particles has been proposed in \cite{lode2012}. 

\begin{figure}
\includegraphics[width=1\linewidth]{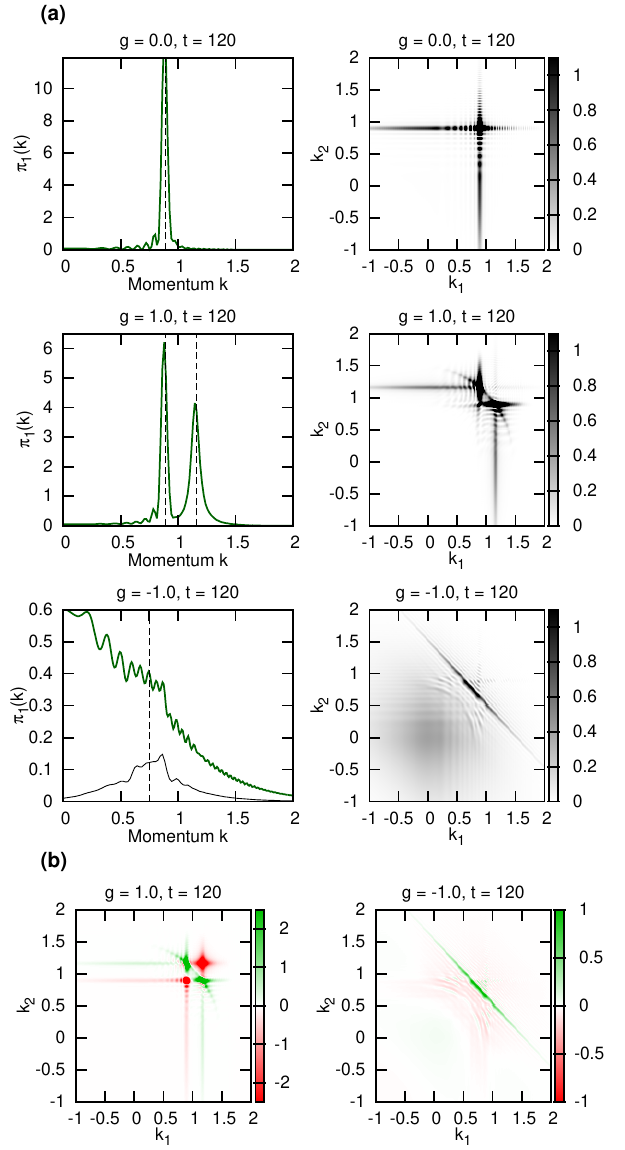}
\caption{(a) The single-particle momentum distribution $\pi_1(k;t)$ and the two-particle momentum distribution $\pi_2(k_1,k_2;t)$ of the two-boson system for various interaction strengths $g$, at a specific moment $t = 120$. Black dashed lines indicate predicted values of characteristic momenta calculated from the system energy (see text). In the non-interacting system ($g = 0.0$) two bosons are emitted sequentially with identical momenta. In the repulsive system ($g = 1.0$) two bosons are emitted sequentially with two different momenta. In a system with sufficiently strongly attractions ($g = -1.0$) the bosons are emitted as a bound pair with a well-defined center-of-mass momentum. For $g = -1.0$, the thin black line shows $\pi_1$ when Eqs. \eqref{MomentumDef} are redefined to exclude the part of the wave function corresponding to the trapped particles. (b) The noise correlation $\mathcal{G}(k_1,k_2;t)$ for the two-boson system at $g = \pm 1.0$ and $t = 120$. The correlations and anticorrelations that cannot be captured by a single-particle description become clearly visible. Momenta are in units of $\sqrt{\hbar m\Omega_0}$, interaction strength in units of $\sqrt{\hbar^3\Omega_0/m}$, time in units of $1/\Omega_0$.}
 \label{Fig5} 
\end{figure}

Initially ($t = 0$) both momentum distributions have nearly-Gaussian shapes centered at $k_1 = 0, k_2 = 0$. For larger times, depending on the tunneling mechanism dominating the dynamics, characteristic features in $\pi_2$ and $\pi_1$ emerge. In Fig.~\ref{Fig5}a we show the single-particle and two-particle momentum distributions for a few different interaction strengths, after the bosons have been allowed to tunnel for some time $t$. To increase understanding of the results, we show these distributions for the same interactions and time moments as in Fig.~\ref{Fig4}.

In the simplest, non-interacting case ($g = 0$), as the particles tunnel from the well, a narrow peak appears in the distribution $\pi_1$, centered around the value $k_0 \approx 0.89$. It is clear that the two bosons are emitted with a very well-defined momentum. The two-particle momentum distribution is a simple product of two identical single-particle distributions $\pi_2(k_1,k_2;t)=\pi_1(k_1;t)\pi_1(k_2;t)$ and clear horizontal/vertical lines at $k_1 = k_0$ and $k_2 = k_0$ are visible. They indicate that the emitted boson has a narrowly defined momentum, while the trapped boson still has a nearly-Gaussian distribution of momenta. 

The momentum characteristics become more complicated for interacting systems. In the case of repulsive interactions ($g = +1.0$, second row in Fig.~\ref{Fig5}a) the dynamics is dominated by the sequential tunneling of bosons. This behavior is reflected in the single-particle momentum distribution $\pi_1(k;t)$ by two distinct peaks. One of them is centered around the non-interacting value $k_0$ while the second one is shifted to larger momenta $k'\approx 1.15$. These two different momenta can be directly associated with momenta of the sequentially emitted bosons. Due to the repulsive interaction, the first boson which leaves the well carries additional energy and in consequence has a higher momentum $k'$. The second boson no longer feels any interaction and therefore it tunnels with the momentum $k_0$. All this means that the momenta of the emitted particles are also causally correlated, {\it i.e.}, the boson can be emitted with the momentum $k_0$ only if the other has been already emitted with momentum $k'$. This specific time-correlation is directly reflected in the two-particle momentum distribution $\pi_2(k_1,k_2;t)$. It is clearly seen that probability of finding the boson with momentum $k_0$ almost vanishes whenever the remaining boson has momentum different than $k'$. Contrary, probability of finding the boson with momentum $k'$ is associated with almost gaussian distribution of the second boson centered around $k=0$, {\it i.e.}, distribution which is characteristic for the trapped boson in the well. 

Particular values for the momenta of emitted bosons $k_0$ and $k'$ can be easily found from analysis of the relevant system energies. In the case studied ($g=+1.0$) one finds that the initial energy of two confined bosons $E_{\mathrm{INI}}(g)\approx 1.07$, while the ground-state energy of a single boson in the well $E_1\approx 0.4$. These energies correspond to $k'=\sqrt{2(E_{\mathrm{INI}}(g)-E_1)} \approx 1.16$ and $k_0=\sqrt{2E_1} \approx 0.89$, respectively (vertical dashed lines in the left middle and left upper plot in Fig.~\ref{Fig5}a). Our numerical results are in full agreement with this phenomenological analysis. 

Note that the above argument also predicts that, for sufficiently strong attractions, when $E_{\mathrm{INI}}(g) < E_1$, sequential tunneling is strongly suppressed by the energy conservation \cite{lode2014,dobrzyniecki2018}. For example, in the case of $g = -1.0$ (bottom row in Fig.~\ref{Fig5}a) the pair tunneling is the dominating mechanism of the decay. Although the single-particle momentum distribution $\pi_1(k;t)$ is quite broad, a clear correlation between momenta of emitted bosons along the line $k_1+k_2=K=\mathrm{const}$  is visible in the two-particle distribution $\pi_2(k_1,k_2;t)$. This indicates that bosons are emitted simultaneously as a bounded pair with a clearly defined center-of-mass momentum $K$ (in this case $K \approx 1.50$), with the particles oscillating around the center-of-mass with opposite relative momenta. Note that in the distribution $\pi_2(k_1,k_2;t)$ also some additional gaussian background centered around $k_1=k_2=0$ is visible. This part of the momentum distribution reflects the momenta of bosons which still remain in the well (see Fig.~\ref{Fig4} for the corresponding density distribution of the remaining particles). 

It is worth noticing that by using a projector $\mathbf{P} = \theta(x_1-x_B)\theta(x_2-x_B)$, the two-particle wave function $\Psi$ can be written as a sum of two orthogonal wave functions, $\Psi_\mathrm{IN} = \mathbf{P}\Psi$ and $\Psi' = (1-\mathbf{P})\Psi$. Then $\Psi_\mathrm{IN}$ encodes the state of exactly two bosons being inside the well while $\Psi'$ encodes the remaining part of the two-particle system. With this decomposition it is possible to study the momentum distribution of the escaping bosons independently of the state of bosons remaining in the well. This approach corresponds to a simple modification of definitions \eqref{MomentumDef} by limiting the wave function to the $\Psi'$ part only. This approach is also justified experimentally since it is possible to measure the momenta of the escaping particles only. With this redefinition, the single-particle distribution is significantly modified since the threshold from confined particles is removed (thin black solid line in the left bottom plot in Fig.~\ref{Fig5}a). After this modification a significant enhancement at momentum $K/2$ (half of the center-of-mass momentum) is clearly visible (vertical dashed line in the left bottom plot in Fig.~\ref{Fig5}a). 

The particular value of the center-of-mass momentum $K$ can be predicted with simple phenomenological argumentation. In this case the initial energy of the system $E_{\mathrm{INI}}(g)$ is fully converted to the energy of the emitted interacting pair $E_{\mathrm{pair}}(g)$. The corresponding pair energy, as noted in Section \ref{sec:eigenspectrum}, is approximately $E_{\mathrm{pair}}(g) \approx (K^2-g^2)/4$. Consequently, $K = \sqrt{4E_{\mathrm{INI}}(g)+g^2}$. In the case studied ($g=-1.0$) one finds $E_{\mathrm{pair}}(g) \approx 0.31$ and $K \approx 1.50$, which agrees very well with the momentum distribution obtained with our numerical approach. 

To make this analysis more comprehensive one can discuss inter-particle correlations not only in terms of the two-particle momentum distribution but also via the so-called {\it noise correlation} \cite{altman2004,mathey2008,mathey2009,brandt2017,brandt2018}. This quantity is defined straightforwardly as the difference between the full two-particle distribution and the product of corresponding single-particle distributions: 
\begin{equation}
{\cal G}(k_1,k_2;t) = \pi_2(k_1,k_2;t)-\pi_1(k_1;t)\pi_1(k_2;t).
\end{equation}
Phenomenologically, the noise correlation can be interpreted as a distribution of correlations which are forced by inter-particle interactions that cannot be captured by any single-particle description. In Fig.~\ref{Fig5}b we plot the noise correlation for two different interactions corresponding to the dominance of two different decay channels ($g=\pm 1.0$). It is evident that a single-particle description strongly underestimates probabilities of finding particles in cases when two-particle momentum distributions display strong correlations (green areas). More importantly, the noise correlation nicely exposes the aforementioned causal correlations between sequentially emitted particles (vertical/horizontal lines localized around $k'\approx 0.89$ for $g=+1.0$).  

\section{The transition}
\label{sec:transition}
The specific transition between different tunneling channels can be analyzed and well described when the momentum of the center of mass $K=k_1+k_2$ is considered. Its distribution can be extracted from the two-particle momentum distribution as follows:
\begin{equation}
	\pi_{\mathrm{CM}}(K;t) = \int\!\mathrm{d}k_2\, \pi_2(K-k_2,k_2;t).
\end{equation}
\begin{figure}
\includegraphics[width=1\linewidth]{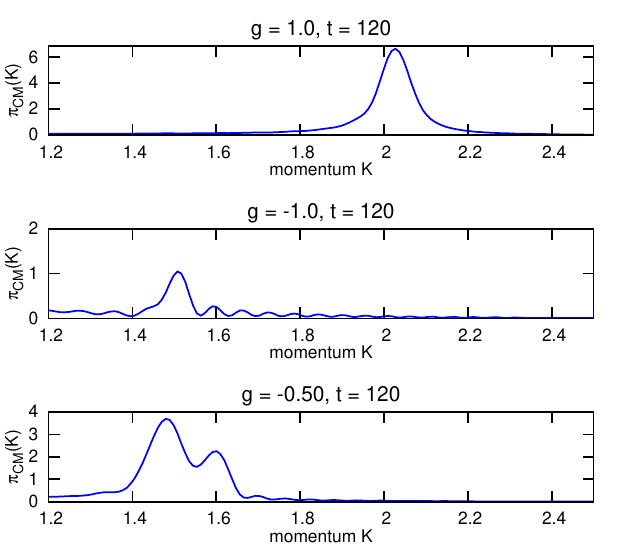}
\caption{The center-of-mass momentum distributions $\pi_\mathrm{CM}(K;t)$ of the two-boson system for various values of $g$, at a specific moment $t = 120$. For the repulsive ($g = 1.0$) and sufficiently strongly attractive ($g = -1.0$) systems, only one decay process is available (sequential and pair tunneling, respectively) and it is reflected in the distribution as a single peak. For a system with weaker attractions ($g = -0.5$) both sequential and pair tunnelings are possible, and two peaks appear in the spectrum, each corresponding to a different process. Momenta are in units of $\sqrt{\hbar m\Omega_0}$, interaction strength in units of $\sqrt{\hbar^3\Omega_0/m}$, time in units of $1/\Omega_0$.}
 \label{Fig6} 
\end{figure}
In Fig.~\ref{Fig6} we display this distribution for three different interactions $g$, after the system has been allowed to evolve for some time $t$. In the case of repulsions ($g=+1.0$) as well as sufficiently strong attractions ($g=-1.0$) a single peak in the center-of-mass momentum emerges. It is centered around the sum of the individual emitted boson momenta $k_0 + k' \approx 2.05$ or the bound pair momentum $K \approx 1.50$, respectively. However, for weaker attractions (for example $g = -0.5$) both tunneling mechanisms are present, and the distribution $\pi_{\mathrm{CM}}(K;t)$ displays two distinct peaks. They can be directly associated with different tunneling processes. By comparing integrated intensities of both peaks it is possible to determine a relative participation of different tunneling mechanisms in the overall dynamics. In Fig.~\ref{Fig7}, we show the relative participation of pair tunneling (green crosses) and sequential tunneling (red triangles) obtained for a few example interaction strengths $g$. For comparison, we include similar quantities (red and green lines) obtained recently by a theoretical analysis of different probability fluxes through the potential barrier \cite{dobrzyniecki2018}. A qualitative agreement between both results opens an additional, much less demanding from the experimental point of view, method for detecting the transition between different tunneling mechanisms. 
\begin{figure}
\includegraphics[width=1\linewidth]{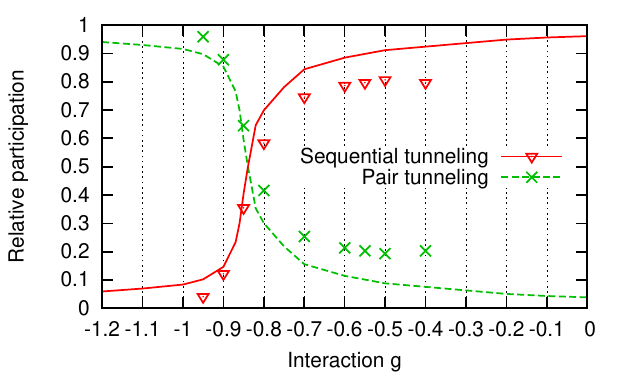}
\caption{The relative participation of pair and sequential tunneling in the overall dynamics of the two-boson system, for various interaction strengths $g$. Green and red symbols show the participation of pair tunneling and sequential tunneling, respectively, calculated from the areas of the corresponding peaks in the center-of-mass momentum distribution $\pi_\mathrm{CM}$ at $t = 180$. For comparison, the corresponding results from Ref. \cite{dobrzyniecki2018} are shown as green dashed and red solid lines, respectively. It can be seen that sequential tunneling dominates in a wide range of interaction strengths, but its participation falls abruptly to zero as $g$ approaches the critical value $g = -0.9$. Interaction strength is given in units of $\sqrt{\hbar^3\Omega_0/m}$.}
 \label{Fig7} 
\end{figure}

Dynamical properties of the system in the vicinity of the transition are also encoded in the interaction energy between particles ${\cal I}(t) = g \int\!\mathrm{d}x\,|\Psi(x,x;t)|^2.$ It is worth noticing that interaction energy is closely related to the Tan's contact $\mathcal{C}$ \cite{rizzi2018}
\begin{equation}
	{\cal C}(t) = \frac{m^2 g^2}{\pi \hbar^4} \frac{\partial {\cal E}}{\partial g}=\frac{m^2 g}{\pi\hbar^4}\,{\cal I}(t)
\end{equation}
which is known to be a very universal quantity linking many different features of atomic systems such as the dependency of energy on interaction strength, the pair correlation function, and the relation between pressure and energy density \cite{yao2018}. Furthermore, it is accessible to experimental measurements \cite{wild2012,bardon2014}. Note that, although the total energy of the system ${\cal E}$ is conserved, its derivative with respect to $g$ changes during the evolution due to the dynamical changes of the system's wave function. 

\begin{figure}
\includegraphics[width=1\linewidth]{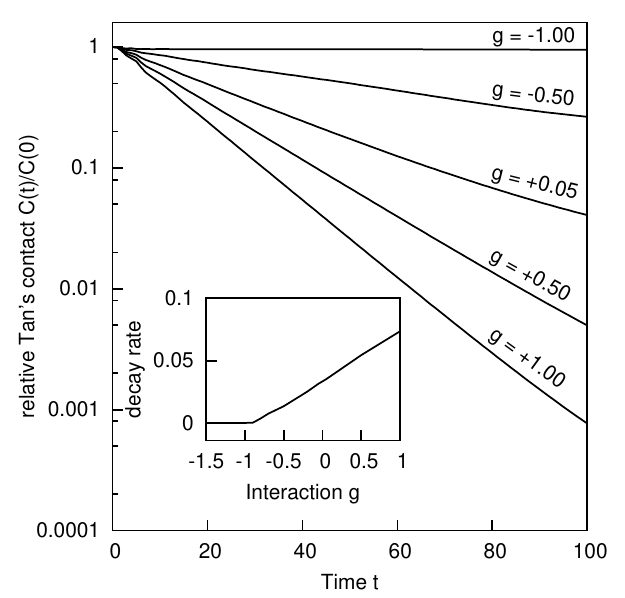}
\caption{Time evolution of Tan's contact relative to its initial value, $\mathcal{C}(t)/\mathcal{C}(0)$, for the two-boson system for various values of $g$. The magnitude of the contact is seen to decay exponentially within the shown timeframe. Inset: The fitted decay rate of the exponential decay of $\mathcal{C}(t)/\mathcal{C}(0)$, as a function of $g$. Time is given in units of $1/\Omega_0$, interaction strength in units of $\sqrt{\hbar^3\Omega_0/m}$, decay rate in units of $\Omega_0$.}
 \label{Fig8} 
\end{figure}

In Fig.~\ref{Fig8} we plot the time evolution of the Tan's contact (relative to its initial value at $t=0$) for different interaction strengths. As it is seen, the contact displays exponential decay. Moreover, the decay rate (inset in Fig.~\ref{Fig8}) strongly depends on the interaction strength and near the transition between sequential and pair tunneling channel at $g \approx -0.9$ it approaches $0$. 

These results can be explained intuitively when the different decay processes are considered. One suspects that the interaction energy, due to the short-range form of interactions, rapidly decreases when particles are sequentially emitted from the trap. Accordingly, when sequential tunneling dominates the dynamics, the magnitude of the contact is closely tied to the probability that the system remains in the initial trapped state. For systems such as the one under study, this probability obeys an exponential decay law to a very good approximation \cite{davydov1976}, hence $\mathcal{C}(t)$ decays exponentially. 

The rate of this exponential decay decreases as the interaction energy in the trapped system is lowered \cite{zurn2012}. Additionally, as the interactions become more attractive, the system dynamics is increasingly dominated by the process of pair tunneling. During pair tunneling the interaction energy remains almost unchanged since particles in all stages of the evolution form a bound pair. Hence, for interaction strengths $g < -0.9$ for which the bosons tunnel only as bound pairs, the decay rate of Tan's contact remains close to zero.

\section{The three-boson case}
\label{sec:3bosons}

Let us now apply the above approach to the system of $N = 3$ bosons. In this case the dynamics is more complicated, as now the bosons can tunnel in more ways: as single particles, as bound pairs, or as bound trimers. Accordingly, the momentum distributions of the system display more complex structures. In order to analyze them, we first define the three-particle momentum distribution $\pi_3(k_1,k_2,k_3;t)$ as
\begin{align}
 \pi_3(k_1,k_2,k_3;t) = \frac{1}{8\pi^3\hbar^3}\left|\int\!\mathrm{d}x_1 \mathrm{d}x_2\,\mathrm{d}x_3 \times \right. \\ \nonumber
  \vphantom{\int_1^2} \left. e^{-i(k_1x_1+k_2x_2+k_3x_3)/\hbar} \Psi(x_1,x_2,x_3;t)\right|^2.
\end{align}

Consequently, the single-particle momentum and center-of-mass momentum distributions $\pi_1,\pi_\mathrm{CM}$ are now defined as
\begin{subequations}
\begin{align}
\pi_1(k;t)&=\int\!\mathrm{d}k' \mathrm{d}k''\,\pi_3(k,k',k'';t),
\\
\pi_\mathrm{CM}(K;t)&=\int\!\mathrm{d}k' \mathrm{d}k''\,\pi_3(K-k'-k'',k',k'';t).
\end{align}
\end{subequations}

\begin{figure}
\includegraphics[width=1\linewidth]{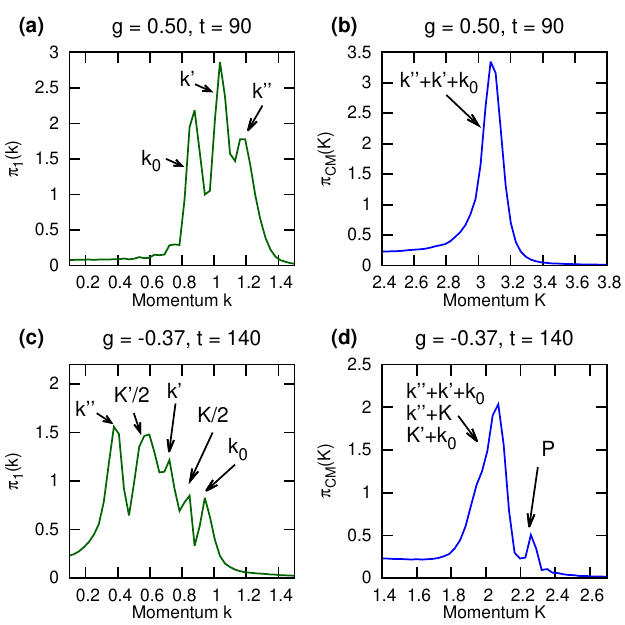}
\caption{The single-particle momentum distribution $\pi_1(k;t)$ and the center-of-mass momentum distribution $\pi_\mathrm{CM}(K;t)$ of the three-boson system for two different interaction strengths, at specific moments $t$. In the repulsive case (top row) three bosons are emitted sequentially with well-defined momenta $k'',k',k_0$. In the attractive case (bottom row) the bosons can additionally tunnel as bound pairs with well-defined center-of-mass momenta $K$ or $K'$, or as a trimer with center-of-mass momentum $P$. Each peak in the distributions can be associated with specific characteristic momenta, as indicated by arrows. Momenta are in units of $\sqrt{\hbar m\Omega_0}$, interaction strength in units of $\sqrt{\hbar^3\Omega_0/m}$, time in units of $1/\Omega_0$.}
 \label{Fig9} 
\end{figure}

In Fig.~\ref{Fig9} we show the distributions $\pi_1$ and $\pi_\mathrm{CM}$ for a three-boson system with different interaction strengths, after the system has been allowed to tunnel for some time $t$. First we focus on the case of repulsive interactions $g = 0.50$. In this case the bosons cannot form bound states and consequently tunnel one at a time, with the successive emitted bosons having momenta $k''$, $k'$ and $k_0$. Much like in the two-boson case, one can obtain these momenta straightforwardly from the relevant system energies. For $g = 0.50$ we find $k'' \approx 1.18, k' \approx 1.05, k_0 \approx 0.89$. This result is directly reflected by the distribution $\pi_1$, which displays three distinct peaks centered near these values (Fig.~\ref{Fig9}a). Similarly, the distribution $\pi_\mathrm{CM}$ displays a single clear peak centered at $k''+k'+k_0 \approx 3.12$, confirming that sequential tunneling is the only available process (Fig.~\ref{Fig9}b). Note that, compared to the two-boson case (Fig.~\ref{Fig5}a), the peaks in Fig.~\ref{Fig9}a are not as well resolved. The main reason is that the characteristic momenta $k'',k',k_0$ fall quite close to each other and therefore the corresponding momentum peaks, having their natural width, are partially overlapping.

The situation changes significantly in the case of weaker attractions (for example $g = -0.37$). In this case the system exhibits the full variety of three-boson tunneling processes, and it can decay in several distinct ways. The first scenario is a sequential tunneling of three independent particles, as described above. In the second scenario, the emission of an independent boson with momentum $k''$ is followed by the emission of a bound pair with center-of-mass momentum $K$. In the third scenario, the first two particles tunnel as a bound pair with center-of-mass momentum $K'$, followed by the remaining particle tunneling independently with momentum $k_0$. The final possibility is that all three bosons tunnel as a bound trimer with center-of-mass momentum $P$. Similarly to the two-boson case, it is straightforward to obtain all these characteristic momenta by analyzing the energy of each emitted group of particles. For $g = -0.37$ we obtain the following values: $k'' \approx 0.42, k' \approx 0.71, k_0 \approx 0.89, K \approx 1.66, K' \approx 1.22, P \approx 2.30$. 

Each of the distinct scenarios of decay is reflected directly in the momentum distributions (Fig.~\ref{Fig9}c and Fig.~\ref{Fig9}d). In the single-particle momentum distribution $\pi_1$, peaks are clearly visible at positions $k'',k',k_0$ corresponding to the sequential tunneling, as well as $K/2$ and $K'/2$ that correspond to the pair tunneling. The distribution $\pi_\mathrm{CM}$ displays two distinct peaks. The first, smaller peak is associated with the trimer tunneling and it is centered near the value $P$. The second, larger peak in $\pi_\mathrm{CM}$ accounts collectively for the single-particle and pair tunneling processes, since their corresponding total center-of-mass momenta fall very close to each other. It can be seen that the distributions $\pi_1$ and $\pi_\mathrm{CM}$, when considered together, can provide information about the complete variety of the participating tunneling processes. 

\begin{figure}
\includegraphics[width=1\linewidth]{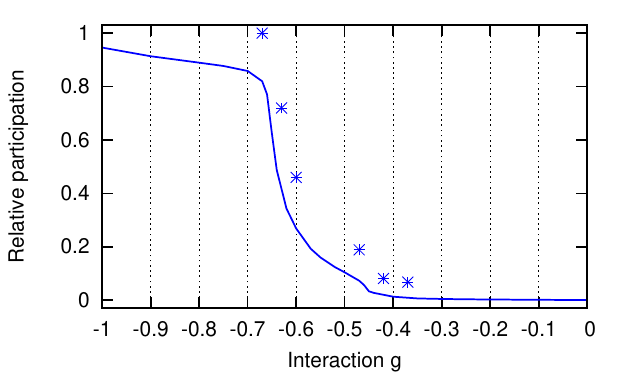}
\caption{The relative participation of trimer tunneling in the overall dynamics of the three-boson system, for various interaction strengths $g$. Blue symbols show the participation of trimer tunneling calculated from the areas of the corresponding peaks in the center-of-mass momentum distribution $\pi_\mathrm{CM}$ at $t = 140$. For comparison, the corresponding result from Ref. \cite{dobrzyniecki2018} is shown as a blue solid line. It can be seen that the participation of trimer tunneling remains near zero for approximately $g > -0.46$, increases throughout the region $-0.65 < g < -0.46$, and becomes close to one for $g < -0.65$. Interaction strength is given in units of $\sqrt{\hbar^3\Omega_0/m}$.}
 \label{Fig10} 
\end{figure}

Analogously to the two-boson system, the three-boson system undergoes a transition between several distinct regimes as the interaction strength is tuned across critical values \cite{dobrzyniecki2018}. These regimes can be identified by analyzing the changing participation of different tunneling mechanisms in the overall decay process. Similarly as in the two-boson case, the proportional participation of the trimer tunneling can be obtained by comparing the integrated intensities of peaks in the distribution $\pi_\mathrm{CM}(K;t)$. In Fig.~\ref{Fig10} we show the relative participation of trimer tunneling obtained by this method for various $g$ (star symbols). For comparison, we also show the analogous quantity (solid line) obtained in \cite{dobrzyniecki2018} from analysis of the probability flux through the potential barrier. There is a qualitative agreement between both results, supporting the validity of the described method for detecting the transition between tunneling mechanisms.

\section{Conclusion}
\label{sec:conclusion}

We have analyzed the decay of a system of a few ultra-cold bosons, initially trapped in an open one-dimensional potential well. In particular, we have examined the influence of the interaction strength $g$ on the dynamics of the momentum distributions of the system, as well as the Tan's contact. We find that there is a essential difference in the behavior of these quantities when the interaction strength $g$ is tuned across a critical value that corresponds to strong suppression of sequential tunneling. These findings are in full agreement with previous results based on careful analysis of many-body probability fluxes \cite{dobrzyniecki2018}. We show that it is possible to establish a relationship between the dominant tunneling process and the form of the momentum distributions. In particular, from the center-of-mass momentum distribution of the system one can quantitatively determine the relative participation of the different tunneling processes in the dynamics. These findings are shown to apply equally well both to $N = 2$ and $N = 3$ systems. Additionally, we examine the evolution of the Tan's contact and show that its behavior also reflects the dominant tunneling process. 

Since the single- and two-particle momentum distributions as well as the Tan's contact are accessible to experimental measurements, the presented results have potential significance for upcoming experiments with ultra-cold bosons in quasi-one-dimensional potentials. The theoretical and experimental analysis of these quantities can give increased insight into the system dynamics. 

Although the results presented here focus on comparably small values of $g$, we have also performed corresponding simulations for the system being in the Tonks-Girardeau regime, \emph{i.e.}, in the limit of infinite repulsive interactions \cite{girardeau1960,kinoshita2004,paredes2004}. The results indicate that the features described in this paper for smaller repulsive interactions remain in effect even for very strong repulsions that approach the Tonks-Girardeau limit.

\section{Acknowledgments}

This work was supported by the (Polish) National Science Center Grant No. 2016/22/E/ST2/00555. Numerical calculations were partially carried out in the Interdisciplinary Centre for Mathematical and Computational Modelling, University of Warsaw (ICM), under Computational Grant No. G75-6.

\end{document}